\documentstyle[prd,aps]{revtex}
\begin{document}
\draft

\title
{Casimir energy for spherical boundaries}

\author {C. R. Hagen\cite{Hagen}}

\address
{Department of Physics and Astronomy\\
University of Rochester\\
Rochester, N.Y. 14627}

\maketitle

\begin{abstract}
     Calculations of the Casimir energy for spherical geometries which are 
based on integrations of the stress tensor are critically examined.  It is
shown that despite their apparent agreement with numerical results obtained 
from mode summation methods, they contain a number of serious errors. 
Specifically, these include (1) an improper application of the stress tensor to
spherical boundaries, (2) the neglect of pole terms in contour integrations,
and (3) the imposition of inappropriate boundary conditions upon the relevant 
propagators.  A calculation which is based on the stress tensor and which 
avoids such problems is shown to be possible.  It is, however, equivalent to 
the mode summation method and does not therefore constitute an independent 
calculation of the Casimir energy. 
\end{abstract}


 
     In 1948 Casimir [1] first predicted that two infinite parallel plates in
vacuum would attract each other.  This remarkable result has its origin in the
zero point energy of the electromagnetic field.  While the latter is highly
divergent, the change associated with this quantity for specific plate 
configurations is found to be finite and thus in principle observable.  Early
work to detect this small effect [2] was characterized by relatively large 
experimental uncertainties which left the issue in some doubt.  More recent
efforts [3] have provided quite remarkable data, but are based on a different 
geometry from that of Casimir.  Since a rigorous theoretical calculation has 
never been carried out for the latter configuration, there remains room for 
skepticism as to whether the Casimir effect is as well established as is
frequently asserted.

     The extension of Casimir's result to problems with nonplanar boundaries
has been of considerable interest and fraught with difficulties.  It was
achieved for the case of the conducting spherical shell by Boyer[4] in a 
remarkable but intricate calculation.  His result was subsequently verified by 
various methods [5-7], including in particular that of direct mode summation 
[8-9].  One of the methods developed in ref. 7 consisted of integrating the 
radial component of the stress tensor over the bounding surface, a technique
which was subsequently applied to the case of the cylinder [10], the circle
[11], and the Dirichlet problem of a D-dimensional sphere [12].  In view of
the increasingly wide application of this technique, it is clearly of interest 
to ascertain its validity.  

Since the electromagnetic sphere is the only known case in which a finite
Casimir energy can be obtained for a nonplanar geometry using conventional 
(e.g., exponential) regularization techniques, it is convenient to use this
case as a specific framework for the present work.  One begins by defining the 
usual stress tensor for the uncoupled electromagnetic field 
\begin{equation}
T^{\mu \nu}=F^{\mu \alpha}F^\nu_{\;\alpha} - {1\over 4}g^{\mu \nu}F^{\alpha
\beta}F_{\alpha \beta}$$
\end{equation}
where $g^{\mu \nu}=(1,1,1,-1)$ and
$$F^{\mu \nu}=\partial^{\mu}A^{\nu}-\partial^{\nu}A^{\mu}$$
with $A^{\mu}$ being the usual vector potential [13].  The latter is taken most 
conveniently to be in the radiation gauge $\partial_kA^k=0, A^0=0$.
The vacuum expectation value of $T^{\mu \nu}$ can clearly be obtained from
appropriate derivatives of the time ordered product
\begin{equation}
G^{ij}({\bf x},t;{\bf x'},t')=i\langle 0|(A^{i}({\bf x},t)A^{j}({\bf
x'},t'))_+|0\rangle
\end{equation}
where the latter is to be evaluated in the presence of spherical conducting
boundaries at $r=a$ and $r=R$ where $r=|{\bf x}|$.  The limit $R\to \infty$ is
to be taken as the final step of the calculation. 

The differential equation 
for $G^{ij}$
$$\left(-\nabla^2 +{\partial^2 \over \partial t^2}\right)
G^{ij}({\bf x},t;{\bf x'},t') = \delta^{ij}({\bf {x-x'}})\delta (t-t')$$
with $\delta^{ij}({\bf x})$ the usual transverse delta function is to be solved
for $0\leq r,r'\leq a$ and for $a\leq r,r' \leq R$ subject to the usual causal 
boundary conditions, namely positive (negative) frequencies for $t-t'>0$ 
($t-t'<0$).  This is an issue of some importance since all previous stress 
tensor calculations of the Casimir energy for nonplanar boundaries have imposed
outgoing wave conditions in the region exterior to the surface of interest.  
While causal boundary conditions follow directly from the existence of the 
vacuum as the state of lowest energy, there is no way to infer outgoing wave 
conditions [14] for the propagator (2).  

The Fourier transform of the propagator is thus inferred from the boundary
conditions to have the form
\begin{equation}
G^{ij}({\bf x}, {\bf x'}; \omega) 
=\sum_n {A^i_n ({\bf x})A^{j*}_n ({\bf x'})\over
\omega_n^2-\omega^2-i\epsilon}
\end{equation}
where the sum is to be taken over all {\em normalized} eigenfunctions $A^i_n$ 
which satisfy the wave equation
$$(\nabla^2 +\omega^2_n){\bf A}_n({\bf r})=0$$
subject to appropriate boundary conditions at $r=0,a,R$ together with the 
transversality condition $\partial_kA^k=0$.  From ${\bf r}\cdot {\bf B}={\bf r}
\times{\bf E}=0$ for $r=a,R$ where $E_k=F^{0k}$ and $B_k=
{1\over 2}\epsilon_{klm}F^{lm}$, 
one readily obtains the form of these functions.  Using
$f_{ln}(\omega^{(1)}_{ln}r)$ and $g_{ln}(\omega^{(2)}_{ln}r)$
respectively to denote the TE and TM modes the eigenfunctions are seen to be of
the form
$${\bf A}^{(1)}_{lmn}({\bf x})=
f_{ln}(\omega_{ln}^{(1)}r){\bf X}_{lm}({\theta, \phi})$$
and
$${\bf A}^{(2)}_{lmn}({\bf x})={1\over \omega_{ln}^{(2)}}
\mbox{\boldmath $\nabla$}\times
g_{ln}(\omega_{ln}^{(2)}r){\bf X}_{lm}(\theta, \phi)$$
where ${\bf X}_{lm}(\theta, \phi)$ with $l\geq 1$ denotes the vector spherical 
harmonics [15], and $f_{ln}$ and $g_{ln}$ are those real linear combinations of
the spherical Bessel functions $j_l$ and $n_l$ which satisfy the boundary 
conditions.  

Denoting the interior and exterior radial functions by $(f^<_{ln},g^<_{ln})$ and
$(f^>_{ln},g^>_{ln})$ respectively, it follows that for $0\leq r\leq a$,
$f^<_{ln}$ and $g^<_{ln}$ are proportional to $j_l$ and required to satisfy the
conditions
$$j_l(\omega_{ln}^{(1)} a)=0$$
$${d\over dr}[rj_l(\omega_{ln}^{(2)}r)]_{r=a}=0$$
from which the interior eigenfrequencies $\omega^{(\lambda)}_{ln}$ are
determined.  For $a\leq r\leq R$ the relevant radial functions are 
$$f^>_{ln}(\omega_{ln}^{(1)}r) =
a_{ln}j_l(\omega_{ln}^{(1)}r)+b_{ln}n_l(\omega_{ln}^{(1)}r)$$
and
$$g^>_{ln}(\omega_{ln}^{(2)}r)
= c_{ln}j_l(\omega_{ln}^{(2)}r)+d_{ln}n_l(\omega_{ln}^{(2)}r)$$
where $a_{ln}$, $b_{ln}$, $c_{ln}$, and $d_{ln}$ are constants. The ratios 
${b_{ln}/a_{ln}}$ and ${d_{ln}/c_{ln}}$ together with the exterior 
eigenfrequencies $\omega_{ln}^{(\lambda)}$ ($\lambda =1,2$) are fixed by the 
vanishing of $f^>_{ln}$ and ${d\over dr}rg^>_{ln}$ at $r=a$ and $r=R$.  This 
enables (3) to be written more explicitly as 
\begin{equation}
G^{ij}({\bf x},{\bf x'}; \omega)=\sum_{lmn\lambda}{A^{(\lambda)i}_{lmn}
({\bf x})A^{(\lambda) j*}_{lmn}({\bf x'})
\over\omega^{(\lambda)2}_{ln}-\omega^2-i\epsilon}
\end{equation}
with ${\bf A}^{(\lambda)}_{lmn}({\bf x})$ normalized according to the 
prescription 
\begin{equation} 
\int d{\bf x}{\bf A}^{(\lambda)}_{lmn}({\bf x})\cdot{\bf A}^{(\lambda ')*}_
{l'm'n'}({\bf x})= 
\delta_{\lambda, \lambda'}\delta_{l,l'}\delta_{m,m'}\delta_{n,n'}
\end{equation}
which in turn is seen to imply the conditions 
$$\int_0^a r^2 dr f^<_{ln} (\omega_{ln}^{(\lambda)}r)^2 =\int_a^R r^2 dr
f^>_{ln} (\omega_{ln}^{(\lambda)}r)^2 = 1$$
and correspondingly for $g^<_{ln}$ and $g^>_{ln}$.  It is to be noted that (4)
applies both to the interior and exterior domains provided only that the
appropriate eigenfrequencies $\omega_{ln}^{(\lambda)}$ and eigenfunctions 
${\bf A}^{(\lambda)}_{lmn}$ are applied in each case.

Having determined the relevant propagators the problem of computing the Casimir
energy $E_c$ can now be addressed.  Using the energy density given in (1), the 
propagators (4), and the normalization conditions (5), the appropriate
derivatives can be taken together with the limits ${\bf x'}\to{\bf x}$ and 
$t'\to t$.  This readily yields the result 
\begin{eqnarray*}
E_c & = & \int_0^R r^2 dr \int d\Omega\langle0|{1\over 2}(E^2+B^2)|0\rangle\\
& = & \sum_{ln\lambda}\left( l+{1\over 2} \right)\omega_{ln}^{(\lambda)}
\end{eqnarray*}
where the summation is to be taken over both the interior and exterior 
eigenmodes of the system.  This sum has been carefully evaluated in ref.9 using
an exponential cutoff with results entirely consistent with previous 
calculations.  Thus it only remains to be determined whether similar 
conclusions follow from the stress tensor method. 
                                                                             
To pursue this issue it is to be noted that the interpretation of 
$T^{kl}(\bf x)$ as the $k$ component of the force per unit area normal to a 
surface $\sigma_l$ comes from the $k$ component of the conservation law

$$\partial_\nu T^{\mu\nu}=0.$$
Thus 

$${\partial\over \partial t}\int_V d{\bf x}\;T^{k0}
= - \int_\Sigma d\sigma_l\; T^{kl}$$
where the integration on the rhs is to be taken over the surface $\Sigma$ which
bounds the volume $V$.  Since the lhs is the time rate of change of the $k$
component of the momentum,  it can then be claimed that $T^{kl}({\bf x})$ is 
the correct force per unit area {\em so long as the calculation is performed 
in cartesian coordinates}.  However, the claim has been made [7,10-12] that it 
is also valid in spherical coordinates, i.e., that the force per unit area
$F/A$ on the spherical surface is simply given by 
\begin{equation}
F/A= \langle0|[T_{rr}(r=a-\epsilon)-T_{rr}(r=a+\epsilon)]|0\rangle.
\end{equation}
This, of course, is an assertion which can be examined by direct calculation, a
task to which attention is now directed. 

At $r=a$ the vacuum expectation value of $T_{rr}$ can be written as 
$\langle0|{1\over 2}(B^2-E^2)|0\rangle$.  
Application of the boundary conditions at $r=a$ 
together with the identity 
$$\sum_m|Y_{lm}(\theta,\phi) |^2={2l+1\over 4\pi}$$
yields the vacuum expectation value of $T_{rr}(r=a-\epsilon)$ as
the angular independent form
\begin{eqnarray}
 &  & \langle 0|T_{rr}(r=a- \epsilon) |0\rangle
  = \sum_{ln}{l+{1\over 2}\over8\pi}
\left\{ {1\over\omega^{(1)}_{ln}} \right. \nonumber \\
& & \left.\left[{d\over dr}f_{ln}^<(\omega^{(1)}_{ln}a)\right]^2
 + \left[\omega^{(2)}_{ln} - {l(l+1) \over \omega^{(2)}_{ln}a^2} 
\right] g_{ln}^< (\omega^{(2)}_{ln}a)^2 \right\}
\end{eqnarray}
where $\omega^{(\lambda)}_{ln}$ refers to the internal eigenfrequencies.  The
corresponding result for 
$\langle 0|T_{rr}(r=a+\epsilon)|0\rangle$ is simply obtained from
(6) by using the corresponding external eigenmodes together with the
replacements $f^<\to f^>$ and $g^<\to g^>$.  In order to relate this result to
the mode summation method, however, an explicit construction of the relevant
eigenfunctions is required.  For the interior region $r\leq a$ the 
normalization of the functions $f_{ln}^<$ and $g_{ln}^<$ leads to
$$f_{ln}^< (\omega_{ln}^{(1)}r) = 
\left\{ {a^3 \over 2\omega_{ln}^{(1)\;2}}
\left[{d\over dr}j_l(\omega_{ln}^{(1)}a)\right]^2
\right\}^{-{1\over 2}}j_l(\omega_{ln}^{(1)}r)$$
and
$$g_{ln}^<(\omega_{ln}^{(2)}r)=\left\{{a\over2}
\left[a^2-{l(l+1)\over \omega_{ln}^{(2)\;2}}\right]j_l
(\omega_{ln}^{(2)}a)^2\right\}^{-{1\over 2}}j_l
(\omega_{ln}^{(2)}r).$$

Upon using these results it is found that 
\begin{equation} 
F/A|_{r = a-\epsilon} = 
{1\over a}({1\over 4\pi a^2)} \sum _{ln\lambda}(l+{1\over2})
\omega_{ln}^{(\lambda)}
\end{equation}
with the sum to be performed over the interior eigenfrequencies.  If a
corresponding result were to obtain for $T_{rr}$ over the exterior
of the sphere, it would then follow that 
\begin{eqnarray}
F/A|_{a-\epsilon}+F/A|_{a+\epsilon} & = & {1\over a}({1\over 4\pi
a^2})E_c\nonumber \\
                               & = & -({1\over4\pi a^2})
\partial/ \partial a \; E_c
\end{eqnarray}
in agreement with the claim of [7].  However, for the exterior of the sphere
there is, of course, a contribution only from $r=a$ while the normalization 
depends upon both $a$ and $R$, a circumstance which in fact prevents one from 
obtaining a result analogous to (8).  To see this it is convenient to go to
manifestly normalized forms by the replacements
$$f^>_{ln}(\omega^{(1)}_{ln}r)\to 
\left\{{r^3\over 2\omega_{ln}^{(1)2}} \left[{d\over dr}f^>
_{ln}(\omega^{(1)}r)\right]^2 
\bigg\vert^R_a\right\}^{-{1\over 2}}f^>_{ln}(\omega^{(1)}_{ln}r)$$
and
\begin{eqnarray*}
g^>_{ln} & ( &\omega^{(2)}_{ln}r)\to \nonumber \\
& & \left\{ {r\over 2}\left[ r^2-{l(l+1)\over
\omega_{ln}^{(2)2}}\right] g^>_{ln}(\omega^{(2)}_{ln}r)^2 
\bigg\vert^R_a\right\}^{-{1\over 2}}g^>_{ln}(\omega^{(2)}_{ln}r).
\end{eqnarray*}
Upon using these expressions in the evaluation of 
${\langle 0|T_{rr}(r=a+\epsilon)|0\rangle}$
there results 
$$F/A|_{r=a+\epsilon}= {1\over a}({1\over 4\pi a^2})\sum_{ln}(l+{1\over 2})
[\omega^{(1)}_{ln}\alpha^{(1)}_{ln}+\omega^{(2)}_{ln}\alpha^{(2)}_{ln}] $$
where 
$$\alpha^{(1)}_{ln}=- { r^3 \left[{d\over dr}f_{ln}^>(\omega_{ln}^{(1)}r)
\right]^2 \big\vert_{r=a}\over r^3 \left[{d\over
dr}f_{ln}^>(\omega_{ln}^{(1)}r)\right]^2\big\vert^R_a }$$
and 
$$\alpha^{(2)}_{ln} = - { r\left[ r^2- {l(l+1)\over\omega^{(2)2}_{ln}}
\right] g_{ln}^> (\omega^{(2)}_{ln}r)^2 \bigg\vert_{r=a} \over 
r\left[ r^2- {l(l+1)\over\omega^{(2)2}_{ln}} \right] g_{ln}^>
(\omega^{(2)}_{ln}r)^2 \bigg\vert_a^R }.$$
Since the coefficients $\alpha_{ln}^{(1,2)}$ vanish in the large $R$ limit, it
follows that this implies a suppression of the contribution of the external
modes in the calculation of the Casimir energy,thereby contradicting Eq.(8).  
Thus the stress tensor approach does not constitute a valid approach to this 
problem for the case of spherical boundaries.  Similar conclusions are readily 
obtained for Dirichlet and Neumann boundary conditions.

Although the conclusions reached here have been based on a very specific
calculation, they can be placed in a more general context by using techniques
of covariant differentiation.  To this end one takes the conservation law
for $T^{\mu\nu}$ in the form
$$T^{\mu\nu}_{,\nu}=0$$
or 
\begin{equation}
\partial _{\nu}T^{\mu\nu} + 
\left\{ \mu\atop \alpha\>\nu \right\} T^{\alpha\nu}
+ \left\{ \nu\atop \alpha\>\nu \right\} T^{\mu\alpha} = 0.
\end{equation}
For spherical coordinates the nonvanishing independent Christoffel symbols are 
$$\left\{\theta\atop r\>\theta\right\}=\left\{\phi\atop r\>\phi\right\}=1/r,$$
$$\left\{r\atop \theta\> \theta\right\}=-r, \qquad
\left\{r\atop\phi\>\phi\right\}=-r^2sin^2\theta,$$ 
$$\left\{\theta\atop \phi\>\phi\right\}=-sin\theta cos\theta, \quad 
{\rm and}\quad \left\{\phi\atop \phi\>\theta\right\}=cot\theta.$$  
This yields for the $r$ component of (10) that
$$\partial_0T^{r0}+{1\over r^3}\partial_rr^3T^{rr} +{1\over
sin\theta}\partial_{\theta}sin\theta
T^{r\theta}+\partial_{\phi}T^{r\phi}={1\over r}T^{00}$$
where use has been made of the tracelessness of $T^{\mu\nu}$.  Since 
$\langle 0|T^{r0}|0\rangle$ 
is time independent and $\langle0|T^{r\theta}|0\rangle=
\langle0|T^{r\phi}|0\rangle=0$, 
it follows that
$$\int dr d\Omega \partial_r(r^3\langle 0|T^{rr}|0\rangle)
= \int r^2 dr d\Omega \langle 0|T^{00}|0\rangle,$$ 
and consequently  
\begin{equation}
r^3\langle 0|T_{rr}(r)|0\rangle \vert^{r_2}_{r_1}
=\int^{r_2}_{r_1}r^2 dr\langle 0|T^{00}(r)|0\rangle
\end{equation}
for arbitrary $r_1$ and $r_2$.  This clearly shows that the correct
inference to be drawn from (10) is not the force equation (6) but rather the
fact that the difference between the quantities 
$r^3\langle 0|T_{rr}(r)|0\rangle$ when 
evaluated on the two bounding spherical surfaces is proportional to the 
total energy in that region.  

It should also be noted that this result can also be inferred directly from
the cartesian tensor result
$$\partial_l\langle0|T^{kl}|0\rangle=0.$$
which implies that
\begin{eqnarray*}
x_k\partial_l\langle 0|T^{kl}|0\rangle 
& = & \partial_lx_k\langle 0|T_{kl}|0\rangle -\langle 0|T^{kk}|0\rangle \\
                               & = &0.
\end{eqnarray*} 
Upon using once again the tracelessness of the energy momentum tensor Eq.(11)
clearly follows. 

It has been noted here that many of the existing calculations of the Casimir 
effect for a sphere have not used correct boundary conditions on the underlying
propagators.  More serious is the fact that those which have used the
stress tensor method have applied an approach which fails in its application to
nonplanar boundaries.  In concluding this work it needs to be pointed out
that there exists yet a third failure in many of these calculations.  To
illustrate this point it is convenient to refer to Eq.(3.14) of ref.[7] which
gives the Casimir energy as 
\begin{eqnarray*}
E_c & = &{i\over 2a}\sum_l(2l+1)\int^{\infty}_{-\infty}{d(\omega a)\over
2\pi}e^{-i\omega \tau}z \nonumber \\
& & \left\{ {(zj_l)'\over zj_l} + {(zj_l)''\over (zj_l)'}+
{(zh_l^{(1)})'\over zh_l^{(1)}} +
{(zh_l^{(1)})''\over (zh_l^{(1)})'} \right\}.
\end{eqnarray*}
where the $j_l$ terms are associated with the interior of the sphere and the
$h_l^{(1)}$ terms with the exterior.  In order to bring this result to the
usual integral over Bessel functions of imaginary argument it is necessary to
perform a ninety degree rotation of the contour.  However, for the case of the
exterior mode part of $E_c$ such a rotation fails since it does not take into 
account the existence of poles of the Hankel functions in the lower half plane. 
This is a noteworthy reminder of the remarks made earlier concerning the fact 
that the outgoing spherical wave condition for $r>a$ cannot be derived for  
Casimir propagators. 

Having displayed some of the problems associated with the calculation of the 
Casimir energy for the case of spherical boundaries it is of interest to note
that calculations which employ the stress tensor method claim to obtain the 
same result as that found by direct mode summation.  This issue is dealt with
in an appendix which shows that when appropriate attention is paid to the issue
of contour rotation one does {\it not} in fact obtain the usual result.
\medskip

\noindent{\bf APPENDIX}

It has been stated in this work that the discontinuity in the stress tensor
across a boundary
cannot be used to calculate the Casimir pressure on that surface. 
Since, however, such calculations invariably
claim to obtain the usual result, it is of 
considerable interest to display explicitly the flaw in such calculations.  In
order to avoid inessential complications one can choose to deal only with
$T_{rr}(r=a+\epsilon)$ for the case of the TE modes with 
corresponding results for
$T_{rr}(r=a-\epsilon)$ and for the TM modes then following immediately. 
This is, of course, equivalent to 
considering the scalar field case with
Dirichlet boundary conditions.  For this case one finds in analogy
to [7] and [12] that the relevant Green's function is obtained by solving the 
equation

$$\left(-\nabla^2-\omega^2\right)G({\bf x,x'};\omega)=\delta({\bf x-x'})$$
subject to Dirichlet boundary conditions at $r=a,R$.  Upon writing 
$$G({\bf x,x'};\omega)=
\sum^{\infty}_{l=0}G_l(r,r';\omega)Y_l^m(\theta,\phi)Y_l^{m*}(\theta ',\phi ')$$
with $Y_l^m$ denoting the usual spherical harmonics, it follows that 
$$G_l(r,r';\omega)=-{1\over r^2W[f^<_{l},
f^>_{l}]}f^<_{l}(\omega r_<)f^>_{l}(\omega r_>)$$
where $W[f^<_{l},f^>_{l}]$ denotes the Wronskian and $r_>$ ($r_<$) denotes the
greater (lesser) of $r$ and $r'$.  In terms of the spherical Bessel functions 
$j_l$ and $n_l$ one readily finds that $f^<_l(\omega r_<)$ and $f^>_l(\omega 
r_>)$ have the form
$$f^<_{l}(\omega r)=j_l(\omega r)n_l(\omega a)-j_l(\omega a)n_l(\omega r)$$
and
$$f^>_{l}(\omega r)=j_l(\omega r)n_l(\omega R)-j_l(\omega R)n_l(\omega r).$$
This allows the partial wave Green's function to be written in the form
$$
G_l(r,r';\omega)={\omega\over n_l(\omega a)j_l(\omega R)-n_l(\omega R)j_l
(\omega a)}f^<_{l}(\omega r_<)f^>_{l}(\omega r_>).
\eqno(A1) $$
Using the fact that the zeros of the Wronskian determine the eigenmodes of the
system $G_l(r,r';t-t')$ is found to be given by
$$ G_l(r,r';t-t')=i\sum_n
{\omega_{ln}f^<_{l}(\omega_{ln} r)f^<_{l}(\omega_{ln} r')\over 2
[{1\over 2}[r^3{d\over dr} f^<_{l}(\omega_{ln} r)]^2|^R_a]}e^{-i\omega |t-t'|}
\eqno(A2)$$
where the sum is taken over all eigenmodes of the system corresponding to
eigenfrequencies $\omega_{ln}$.  This form of the propagator is recognizable as
the Fourier transform of
the scalar version of (3) with explicitly normalized eigenfunctions.  An
immediate consequence is that the calculation of Casimir
energy
by direct mode summation can now be carried out using the approach of [9] to
obtain the usual result.  In addition the analysis following Eq.(9) can be 
repeated, again finding that there is no contribution to $\langle 0|T_{rr}
(r=a+\epsilon)|0\rangle$ in the large $R$ limit.
 
Although this would seem to establish the inapplicability of the stress tensor
method, it is instructive to continue the analysis to 
determine exactly at what point the
latter method fails.  To this end one notes that the evaluation of
$\langle 0|T_{rr}(r=a+\epsilon)|0\rangle$ requires that derivatives with 
respect to both $r$ and $r'$ be taken at $r=r'=a$ and an integration be
performed over $\omega$.  Since the eigenmodes occur at real values of
$\omega$, the path of integration must be carefully specified.  As deduced from
the boundary conditions (and also as stated clearly in [7]) the appropriate
path is just above (below) the real axis for $\omega>0$ ($\omega<0$).  The
usual expression for the Casimir force is then obtained if (a) the contour
can be rotated ninety degrees counterclockwise and (b) if $f^>_l(\omega a))$
becomes proportional to the Hankel function $h^{(1)}_l(\omega a)$ 
($h^{(2)}_l(\omega a))$ 
in the upper (lower) half plane in the large $R$ limit.  
Although the second condition (b) is satisfied, giving a $R$ independent 
result which is formally identical to the usual expression for the Casimir
energy, it is easy to see that the required contour rotation cannot be
performed.  To display this result one writes 
$$ \langle 0|T_{rr}(r=a+\epsilon)|0\rangle =\lim_{t=t'+\epsilon}\int^{\infty}_
{-\infty}d\omega e^{-i\omega (t-t')}
{i\over 2r^2W [f^<_l,f^>_l] }
[{d\over dr}f^<_l(\omega r){d\over dr}f^>_l(\omega r)]|_{r=a}
$$ 
which can be reduced to
$$ \langle 0|T_{rr}(r=a+\epsilon)|0\rangle={-i\over 2a^2}\lim_{t=t'+\epsilon}
\int^{\infty}_{-\infty}d\omega e^{-i\omega (t-t')}{j_l(\omega R){d\over dr}n_l
(\omega r)-n_l(\omega R){d\over dr}j_l(\omega r)\over 
j_l(\omega R)n_l(\omega a) - j_l(\omega a) n_l(\omega R)}\bigg|_{r=a}.
$$ 
While it is clear that the segment of the contour from $\omega=-\infty$ to
$\omega=0$ can be rotated counterclockwise to the negative imaginary axis, no
such rotation is possible for the segment $\omega=0$ to $\omega=\infty$ [16]. 
Conversely, in the case that the stress tensor is defined by the limit $t=t'
-\epsilon$ rather than $t=t'+\epsilon$, it becomes possible to rotate that part
of the contour along the positive real axis to the positive imaginary axis. 
This choice, however, precludes the possibility of a legitimate rotation of the
contour along the negative real axis. This completes the proof that the stress 
tensor does not in fact yield the usual expression for the Casimir stress on 
the sphere.  The demonstration in the body of this paper was based on the 
method of expansion in eigenfunctions and in this appendix on the Wronskian
formulation of the Green's function, the latter being the one more commonly 
used in published calculations of the stress tensor.  The results of the two 
approaches are identical (note the remark [16]) and show that stress tensor 
calculations in curvilinear coordinates cannot be expected to yield the correct
Casimir energy.
 
\acknowledgments

This work is supported in part by the U.S. Department of Energy Grant
No.DE-FG02-91ER40685.


\begin{references}
\bibitem{ } Since the usual view of the Casimir effect sees it as arising
from the zero point energy of the electromagnetic field, it is inappropriate to 
impose such time asymmetric boundary conditions which have the further
consequence of eliminating any possible correspondence to the case of planar
boundaries.
\bibitem{ } It is of interest to note here that if the rotation of the contour
along the negative real axis is continued to just below the positive real axis
a simple application of Cauchy's theorem yields (A2).  The latter is known to
give zero stress in the limit of large $R$ by the analysis following Eq.(9).
\end{references}
\end{document}